\documentstyle[epsfig]{mn}

\newif\ifAMStwofonts

\def\msun{{\rm M_{\odot}}}

\def\be{\begin{equation}}
\def\ee{\end{equation}}

\ifoldfss
  \ifCUPmtlplainloaded \else
    \NewTextAlphabet{textbfit} {cmbxti10} {}
    \NewTextAlphabet{textbfss} {cmssbx10} {}
    \NewMathAlphabet{mathbfit} {cmbxti10} {} 
    \NewMathAlphabet{mathbfss} {cmssbx10} {} 
  \fi
  \ifAMStwofonts
    \ifCUPmtlplainloaded \else
      \NewSymbolFont{upmath} {eurm10}
      \NewSymbolFont{AMSa} {msam10}
      \NewMathSymbol{\upi}     {0}{upmath}{19}
      \NewMathSymbol{\umu}     {0}{upmath}{16}
      \NewMathSymbol{\upartial}{0}{upmath}{40}
      \NewMathSymbol{\leqslant}{3}{AMSa}{36}
      \NewMathSymbol{\geqslant}{3}{AMSa}{3E}

    \fi
  \fi
\fi 

\ifnfssone
  \newmathalphabet{\mathit}
  \addtoversion{normal}{\mathit}{cmr}{m}{it}
  \addtoversion{bold}{\mathit}{cmr}{bx}{it}
  \newmathalphabet{\mathbfit} 
  \addtoversion{normal}{\mathbfit}{cmr}{bx}{it}
  \addtoversion{bold}{\mathbfit}{cmr}{bx}{it}
  \newmathalphabet{\mathbfss} 
  \addtoversion{normal}{\mathbfss}{cmss}{bx}{n}
  \addtoversion{bold}{\mathbfss}{cmss}{bx}{n}
  \ifAMStwofonts
    \ifCUPmtlplainloaded \else
      \UseAMStwoboldmath
      \makeatletter
      \new@mathgroup\upmath@group
      \define@mathgroup\mv@normal\upmath@group{eur}{m}{n}
      \define@mathgroup\mv@bold\upmath@group{eur}{b}{n}
      \edef\UPM{\hexnumber\upmath@group}
      \new@mathgroup\amsa@group
      \define@mathgroup\mv@normal\amsa@group{msa}{m}{n}
      \define@mathgroup\mv@bold\amsa@group{msa}{m}{n}
      \edef\AMSa{\hexnumber\amsa@group}
      \makeatother
      \mathchardef\upi="0\UPM19
      \mathchardef\umu="0\UPM16
      \mathchardef\upartial="0\UPM40
      \mathchardef\leqslant="3\AMSa36
      \mathchardef\geqslant="3\AMSa3E
    \fi
  \fi
\fi 

\ifnfsstwo
  \DeclareMathAlphabet{\mathbfit}{OT1}{cmr}{bx}{it}
  \SetMathAlphabet\mathbfit{bold}{OT1}{cmr}{bx}{it}
  \DeclareMathAlphabet{\mathbfss}{OT1}{cmss}{bx}{n}
  \SetMathAlphabet\mathbfss{bold}{OT1}{cmss}{bx}{n}
  \ifAMStwofonts
    \ifCUPmtlplainloaded \else
      \DeclareSymbolFont{UPM}{U}{eur}{m}{n}
      \SetSymbolFont{UPM}{bold}{U}{eur}{b}{n}
      \DeclareSymbolFont{AMSa}{U}{msa}{m}{n}
      \DeclareMathSymbol{\upi}{0}{UPM}{"19}
      \DeclareMathSymbol{\umu}{0}{UPM}{"16}
      \DeclareMathSymbol{\upartial}{0}{UPM}{"40}
      \DeclareMathSymbol{\leqslant}{3}{AMSa}{"36}
      \DeclareMathSymbol{\geqslant}{3}{AMSa}{"3E}
    \fi
  \fi
\fi 

\ifCUPmtlplainloaded \else
  \ifAMStwofonts \else 
    \def\upi{\pi}
    \def\umu{\mu}
    \def\upartial{\partial}
  \fi
\fi

\title{Superhumps in Low--Mass X--Ray Binaries}
\author[C.A. Haswell, A.R. King, J.R. Murray, P.A. Charles  ]
       {C. A. Haswell,$^1$
       A. R. King,$^2$ J.R. Murray,$^2$
       P. A. Charles$^3$ \\
        $^1$Department of Physics and Astronomy, Open University, Walton Hall,
Milton Keynes, MK7 6AA, UK\\
        $^2$Department of Physics and Astronomy, University of Leicester, 
University Road, Leicester LE1 7RH, UK\\
        $^3$Department of Physics and Astronomy, University of Southampton, 
Hampshire, SO17 1BJ, UK\\
}

\date{Accepted
      Received }

\pagerange{\pageref{firstpage}--\pageref{lastpage}}
\pubyear{2000}

\begin{document}

\maketitle

\label{firstpage}

\begin{abstract}
We propose a mechanism for the superhump modulations observed in
optical photometry of at least two black hole X-ray transients
(SXTs). As in extreme mass--ratio cataclysmic variables (CVs),
superhumps are assumed to result from the presence of the 3:1 orbital
resonance in the accretion disc. This causes the disc to become
non--axisymmetric and precess. However the mechanism for superhump
luminosity variations in low mass X-ray binaries (LMXBs)
must differ from that in CVs, where it
is attributed to a tidally--driven modulation of the disc's viscous
dissipation, varying on the beat between the orbital and disc
precession period. By contrast in LMXBs, tidal dissipation in the outer
accretion disc is negligible: the optical emission is overwhelming
dominated by reprocessing of intercepted central X-rays. Thus a
different origin for the superhump modulation is required. Recent
observations and numerical simulations indicate that in an extreme
mass--ratio system the disc area changes on the superhump period. We
deduce that the superhumps observed in SXTs arise from a modulation of
the reprocessed flux by the changing area. 
Therefore, unlike the situation in CVs, where the
superhump amplitude is inclination--independent, superhumps should be best
seen in low--inclination LMXBs, whereas an orbital modulation
from the heated face of the secondary star should be more prominent at
high inclinations.
Modulation at the disc precession period (10s of
days) may indicate disc asymmetries such as warping. We comment on the
orbital period determinations of LMXBs, and the possibility and
significance of possible permanent superhump LMXBs.

\end{abstract}


\section{Introduction}
Superhumps are periodic optical modulations observed in superoutbursts
of the SU UMa dwarf novae (see Warner, 1995 for a review). Their most
striking property is that their periods $P_{\rm sh}$ are slightly
longer than the orbital period $P_{\rm orb}$, typically by
$1-7$~\%. Since the work of Whitehurst (1988), Whitehurst \& King
(1991) and Lubow (1991a, b) it is now understood that superhumps are a
consequence of the presence of the 3:1 orbital resonance within the
accretion disc. This causes the disc to become eccentric, and to
undergo slow prograde precession in the inertial frame. The secondary
star thus repeats its motion with respect to the disc on the beat
period between the orbit and this precession, which is therefore
slightly longer than $P_{\rm orb}$. This relative motion of the
secondary star modulates the disc's streamline geometry on the superhump
period, causing its viscous dissipation to vary on the same
period. This intrinsic variation of the disc light satisfies one
of the basic observational features of CV superhumps, namely that their
occurrence is independent of binary inclination (Warner 1995, section 3.6.4.1).

The defining resonance condition restricts the mass ratio $q =
M_2/M_1$ to extreme values
\begin{equation}
q \la 0.33.  \label{1}
\end{equation}
Since in most CVs the white dwarf mass $M_1$ lies in a narrow range
($\sim 0.6 - 0.8\msun$), and $M_2$ is often strongly correlated with
the orbital period $P = P_{\rm hr}$~hr (i.e. $M_2/\msun \simeq
0.11P_{\rm hr}$), this confines the occurence of superhumps to short
orbital periods, mostly below the CV period gap (so $P_{\rm hr} \la
2$). In practice almost all of these systems are dwarf novae, with the
superhumps occurring during superoutbursts. However persistent systems
satisfying the resonance condition (\ref{1}) do exist, and show the
superhump phenomenon permanently (`permanent superhumpers', Patterson
1999). `Negative superhumps', in which  $P_{\rm SH} < P_{\rm orb}$,
also exist, and clearly correspond to retrograde disc precession in
the inertial frame. There is no currently accepted explanation
for the retrograde precession inferred from negative superhumps
in CVs. 

Murray (2000) shows that the precession period,
and thus the quantity 
\begin{equation}
\epsilon = {P_{\rm sh} - P_{\rm orb} \over P_{\rm orb}}, 
\label{2}
\end{equation}
depend on both the mass ratio and other conditions such as the disc
pressure or temperature. Small mass ratios $q$ produce small $\epsilon$, 
as does a high disc temperature. 
Taking this to extremes, 
an extreme mass ratio system with a very hot disc could 
have a disc with retrograde precession.

\section{Superhumps in LMXBs}
The resonance condition (1) is easily fulfilled in
binaries with large primary masses $M_1$. The prevalence of superhumps
in outbursting CVs suggests that soft X--ray transients, particularly
those containing black holes, (see Marsh 1998 for a review) might be prime
candidates for showing superhumps. Recognising this, O'Donoghue and
Charles (1996) carefully surveyed the available observational
evidence, and concluded that superhumps had been seen in outbursts of
at least two black--hole SXTs (GS1124--68 = X-Ray Nova Mus 1991, GRO
J0422+32, and probably GS 2000+25). 
%
%
At about the same time,
superhumps were claimed in observations from the 1993 outburst
of GRS 1716-249 (=X-Ray Nova Oph 1993 = GRO J1719-24, Masetti et al 1996),
but as the orbital period of this system is unknown, this claim
cannot be rigorously assessed.
It is notable that the superhump excesses $\epsilon $ are only $\sim 1 - 2
\%$, 
which with $P_{\rm orb} \ga 5 {\rm hr}$,
implies much longer disc precession periods, $P_{\rm prec} =
P_{\rm sh}/\epsilon \sim 10 - 50$~days, than in CVs. 

Although supporting the importance of the 3:1 orbital
resonance, the existence of superhumps in LMXBs presents us with an
apparent difficulty. The intrinsic dissipation in LMXB discs has long
been known to be a negligible contributor to their optical light
(e.g. van Paradijs and McClintock 1995).  The argument is simple: the
X--ray luminosity $L_X$ gives an estimate of the accretion luminosity
$\eta \dot M c^2$ (where $\eta \simeq 10^{20}$~erg~g$^{-1}$ is the
efficiency of rest--mass conversion) and thus the accretion rate $\dot
M$ on to the central star (neutron star or black hole) in an
LMXB. This immediately gives an estimate of the optical luminosity
$L_{\rm opt}({\rm visc})$ of a steady--state accretion disc
surrounding the central object (cf Frank et al., 1992). For all
persistent LMXBs the ratio $L_{\rm opt}({\rm visc})/L_X$ predicted by
this method is far smaller 
than the observed value (van Paradijs \& McClintock 1995). 
This conclusion can be extended to
SXTs in outburst, as their discs can then be regarded as approximately
steady.

The explanation for the excess optical luminosity of LMXBs is
straightforward. If the disc intercepts even a small fraction of the
central X--ray luminosity, this completely dominates its own intrinsic
dissipation.  For a point source at the centre of the disc, the
irradiation temperature $T_{\rm irr}$ is given by
\begin{equation}
T_{\rm irr}^4 = {\eta \dot Mc^2(1-\beta)\over 4\pi \sigma R^2}{H\over R}g.
\label{3}
\end{equation}
(e.g. van Paradijs, 1996)
Here $\beta$ is the X--ray albedo, $H(R)$ is the local disc
scaleheight, and 
\begin{equation}
g = \biggl({{\rm d}\ln H\over {\rm d}\ln R} - 1\biggr). 
\label{4}
\end{equation}
Viscous dissipation alone gives an effective temperature $T_{\rm visc}$, with 
(e.g. Frank et al.\, 1992)
\begin{equation}
T_{\rm visc}^4 = {3GM\dot M \over 8\pi \sigma R^3} 
\label{5}
\end{equation}
at disc radii $R$ much larger than the radius, $r_*$, of the central object  (mass $M$).
Dividing, one finds
\begin{equation}
{T_{\rm irr}^4 \over T_{\rm visc}^4} = {2\eta c^2\over
3GM}(1-\beta)\biggl({H\over R}\biggr) g R
\label{6}
\end{equation}
For LMXBs, the combination $(2\eta c^2/3GM)(1-\beta)$ is of order
$r_*^{-1}$, where $r_*$ is the radius (e.g. event horizon) of the central star.  Thus for
a large enough disc, i.e. one with $R >> r_*(R/Hg)$, irradiation wins
over intrinsic dissipation despite the small solid angle $\sim H/R$ of
the disc, because $T_{\rm irr}^4$ falls off only as $R^{-2}$, whereas
$T_{\rm visc}^4$ goes as $R^{-3}$. This condition is very easily
satisfied in all LMXBs ($R/r_* \ga 10^4$, while $R/Hg \la
10^3$). $L_{\rm opt}$ is thus predominantly a result of disc
irradiation. In agreement with this, van Paradijs \& McClintock (1994)
show that for a sample of 18 LMXBs the observed $L_{\rm opt}$ scales
with $L_X$ and disc size approximately as expected.

It is perhaps worth noting that the mere existence of efficiently
irradiated discs is a challenge to theory, which usually predicts that
a disc heated by central X--rays adopts a convex shape (formally $g <
0$), and thus shields most of its area from the central flux
(e.g. Cannizzo 1994, Dubus et al 1999). However the
observational evidence that LMXB discs {\it are} irradiated is
overwhelming, and we adopt this view here (cf King \& Ritter,
1998).

The overwhelming dominance of irradiation over intrinsic dissipation means
that the explanation of the superhump luminosity variations in CVs
will not work for LMXBs. The intrinsic dissipation at a given disc
radius $R$, e.g. the resonant radius, varies as the local accretion
luminosity $L_{\rm acc}(R) = GM\dot M/R$: the compactness $M/r_*$
of the central object is irrelevant. The intrinsic superhump
luminosities $L_{\rm sh}$ in LMXBs and CVs are thus in the ratio $(M/R)_{\rm
LMXB}/(M/R)_{\rm CV}$. Even for black--hole systems, where $M_{\rm
LMXB}/M_{\rm CV} \sim 10$, the longer orbital periods and
thus larger values of $R$ make this ratio of order unity.  We conclude
that intrinsic LMXB superhumps have luminosities $L_{\rm sh}$ similar
to those in CVs. But the latter are a fraction $f \la 0.1$ only of the total
intrinsic optical disc luminosity $L_{\rm opt}({\rm visc})$, 
and the same will be true in LMXBs. Thus
intrinsic superhump luminosity variations in LMXBs have amplitudes 
\begin{equation}
{L_{\rm sh} \over L_{\rm irr}} = 
{L_{\rm sh} \over L_{\rm opt}({\rm visc})}{L_{\rm opt}({\rm
visc})\over L_{\rm irr}} < f \times 10^{-3} \la 10^{-4}
\label{7}
\end{equation}
of the observed disc brightness resulting from irradiation. Superhump
variations powered by intrinsic viscous dissipation are therefore
negligible in LMXBs, and cannot explain the observed superhump amplitudes.

\section{Disc Area Variations}

The work of the last Section shows that one can only rescue the
resonance theory of superhumps for LMXBs if the efficiency of the
precessing disc in reprocessing the central X--rays varies on the
superhump period. Writing 
\begin{equation}
L_{\rm opt} \propto A_{\rm eff} L_X,
\label{8}
\end{equation}
we see that this requires that the effective area $A_{\rm eff}$ which
the disc presents to the X--rays must vary on this cycle. In principle
this could occur because the disc aspect ratio might vary on this
cycle. 
Smale et al. (1992) suggest this occurs 
because the vertical component of gravity for the disc rim 
will be significantly increased when the outermost part of
the elliptical disc coincides in azimuth with the mass donor
star.  They do not, however, present any quantitative
assessment of how large an effect this will be.

\begin{figure}
\epsfig{width=3.0in,file=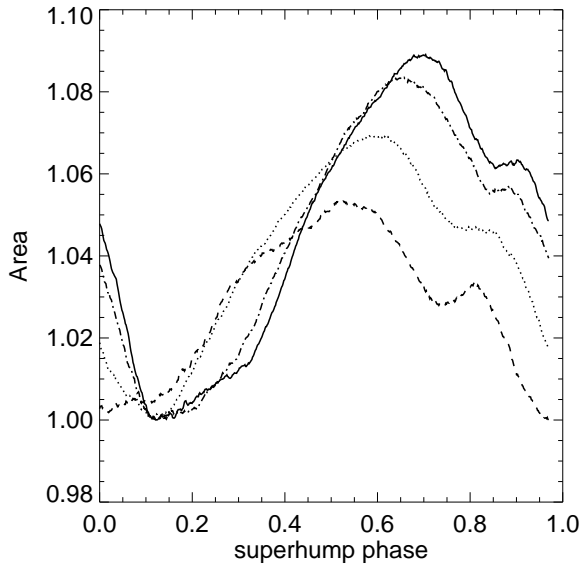}
%

%
\caption
{
Disc surface area as it varies over the course of a single superhump
cycle. The solid line shows the change in surface area of the entire disc
(+ stream), the dot-dash line shows disc region with densities $> 1\%$ 
of the maximum
density, dotted line shows regions with density $> 5\%$ of the
maximum and the dashed
line shows regions with density $> 10\%$ of the maximum.
In this simulation the superhump period is $1.0295 P_{\rm orb}$ .
}
\label{area}
\end{figure}

In contrast,
by far the simplest possibility is that the total disc
surface area varies. This may indeed have been seen in observations of a
persistent superhumper, V348 Pup (Rolfe et al, 2000). Moreover, SPH
simulations also predict a $\sim 10\%$ variation of the
area. Figure~\ref{area} shows the disc area changes found in a
simulation (Murray 2000) of the dwarf nova OY Car. This
simulation assumed a mass ratio $q=0.102$, similar to those measured 
in black hole SXTs, and found a superhump
period excess $\epsilon = 0.0295$.  
The maxima of the area variations are in phase with the intrinsic
viscous dissipation maximum, while the area minima lag the dissipation
minimum by about 45 degrees. From (\ref{8}) we see that the area
variations of Fig.~\ref{area} give approximately the predicted optical
light curve, apart from the effect of dilution by other sources of
optical emission. The most important of these is the X--ray heated
face of the secondary star, whose effective temperature $T_{\rm irr,\
2}$ we can crudely estimate from
\begin{equation}
T_{\rm irr,\ 2}^4 \simeq {\eta \dot Mc^2(1-\beta)\over 4\pi \sigma
a^2}{R_2^2\over 4a^2},
\label{9}
\end{equation}
with $a$, $R_2$ the binary separation and secondary radius. Since
$R_2/a \simeq 0.462[q/(1+q)]^{1/3}$ (Roche geometry), and the disc radius
$R$ is a substantial fraction of $a$, we see from (\ref{3}) that the
heated secondary and the disc have comparable optical brightness. In
an LMXB observed at low inclination this effect will reduce the
predicted fractional superhump amplitude somewhat below the variations
seen in Fig.~\ref{area}. At higher inclinations the changing
aspect of the heated face will produce an optical modulation at
$P_{\rm orb}$ which will compete with the superhump variation at
$P_{\rm sh}$. Moreover the foreshortening of the disc at such
inclinations will weaken the superhump modulation. We would therefore
expect to see both modulations only in rare cases, and SXTs to divide
into groups showing one or other of them. The similarity of (\ref{9})
and (\ref{3}) suggests that these two groups should contain roughly
equal numbers of systems, approximately as observed (see below).
The neutron star transient XTE J2123-058 is at high orbital inclination,
and showed a clear optical modulation at $P_{\rm orb}$, comprised of
the changing aspect of the heated face and grazing eclipses (Zurita et al 
2000).  There was no trace of a superhump modulation, but this
is to be expected for a neutron star system with $P_{\rm orb}=6.0~{\rm
hr}$, since the mass ratio is unlikely to satisfy (1). 

\section{Discussion}
We have shown that variations of the disc surface area on the
superhump cycle offer a plausible explanation for the superhump light
curves observed in LMXBs. 
The
superhump modulation in area (Figure 1) shows
that the geometry
of the accretion flow varies on the superhump period.
Therefore: 
(i) the area of the disc visible to the
observer 
changes,
causing a modulation in  
the optical flux;
(ii) the solid angle the disc subtends at the X-ray source will probably
change,
causing 
a modulation in the intercepted fraction of $L_{\rm X}$.
Both these factors contribute to the superhump modulation.

The small observed values of $\epsilon$ are
a natural consequence of the extreme mass ratios $q$ and high disc
temperatures in LMXBs. The incidence of detected superhump variations,
currently unambiguously seen in 2 or 3 
SXTs, is in line with expectations from
this model, as data able to address the issue was not
collected for most SXTs. 
If the disc had any front--back asymmetry, perhaps as a result
of warping under radiation--induced torques 
(Pringle 1996, Wijers and Pringle 1999),
detection of optical
modulation on the disc precession period $P_{\rm prec}$ might be possible. 
However
the small observed $\epsilon$ values mean that $P_{\rm prec}$ is of
order 10s of days, and thus frequently comparable with the duration of
the outburst itself. Since neither $P_{\rm orb}$ nor $P_{\rm prec}$ will 
in general be known
until the outburst has finished, this highlights the importance of
obtaining accurate photometry of any `orbital' variations at as many
stages of an outburst as possible.

In SXTs we can establish $P_{\rm orb}$ accurately from radial velocity
measurements in quiescence, and thus confidently say whether a given
photometric modulation is or is not a superhump. In a persistent LMXB,
it is generally only possible to determine 
$P_{\rm orb}$ if the system is at high orbital inclination so
that X-ray eclipses or dips are seen as the central source is
periodically occulted by the mass donor star or vertical structures
in the disc rim.
Of the five possible LMXB superhumpers listed in Ritter and Kolb (1998)
four are SXTs. The fifth is the dipping source V1405 Aql 
which
was originally suggested as a possible superhumper by White (1989).
We discuss this source and GX\thinspace 9+9 individually, before
giving a general discussion.

\subsection{V1405 Aql (A\thinspace 1916-05, 4U\thinspace 1915-05)}

V1405 Aql was 
comprehensively studied by Callanan, Grindlay and Cool (1995, hereafter CGC).
Their observational findings were
\begin{enumerate}
\item
the optical period is 50.4589 mins
\item
the X-ray period is 50.00 mins
\item
in an ``anomalously low optical state'' (0.5 mag below normal)
         both 50.46 and 50.00 min periods were observed in the optical
\item
the 50.4589 min period seems to have phase stability over 7 years.
\end{enumerate}
CGC's conclusions were first
that 50.4589 mins is 
the orbital period
$P_{\rm orb}$,
since the superhumps in outbursting SU UMa stars show neither period nor 
phase stability over weeks, let alone years.  Second, 
in 
the anomalously low optical state the intrinsic (i.e. viscously generated) 
brightness
      of the disc should make a significant contribution, rendering ``SU
      UMa - like precessing behaviour'' more observable. 
Finally
phase-wandering of the optical dips when folded on the X-ray period
prevents
simple interpretation of the periodicities.

Since 1915-05 is a steady X-ray source, it is misleading
to compare it with the
SU UMa CVs in which the disc is subject to the thermal-tidal instability.
SU UMa stars exhibit superhumps during their
outbursts, when their discs are in the high viscosity state for 
days to weeks. 
SU UMa discs therefore 
change their radii and mass distributions
on timescales of days, and their superhumps
evolve as a consequence of this.  For a steady disc satisfying equation (1)
we should expect the superhumps to settle into a stable oscillation:
simulations of precessing eccentric discs settle into periodic
behaviour unless the boundary conditions change (Murray 1998, 
Whitehurst private communication). 
We 
conclude
that the phase stability of the 50.4589 min period
is
no barrier to interpreting it as ${P_{\rm sh}}$.
We accordingly identify the 50.00 min period as $P_{\rm orb}$.

The ratio of 
viscously--generated to irradiation--generated flux remains constant 
(equation~\ref{6}) unless either the disc is 
gaining or losing
mass at some radius, or the geometry changes.  In fact all the
non-negligible components of the optical light curve: $L_{\rm acc}$, $L_{\rm irr}$, 
and the irradiated flux from the mass donor, remain in the same ratios
for a steady state accretion flow with any $\dot M$ unless the geometry changes.
Therefore it is likely that CGC's ``anomalously low optical state''
coincides with
a change in the geometry of the accretion flow, making
irradiation of the mass donor star more prominent. If
the geometry did not change there would be no reason for the modulation
at $P_{\rm orb}$ to become more prominent (the ellipsoidal variation
would introduce a signal at $P_{\rm orb}/2$, but is never
likely to be observable as the intrinsic flux from the mass donor
is certain to be negligible in a short period X--ray--emitting 
neutron star binary like V1405 Aql).

Since we are proposing that the geometry of the disc changes on
$P_{\rm sh}$,
the deep dips seen in figure 7 of CGC could well
be analogous to the ``superdip'' seen in OY Car by Billington
et al (1996).   
This interpretation
can also explain the phase-wandering of the optical dips when
they are folded on $P_{\rm orb} = 50.00$~mins, hence negating
the third of CGC's conclusions.

We identify V1405 Aql as a persistent irradiated superhumper.
With our interpretation $\epsilon = 0.009$,
 and the inferred period of apsidal disc
precession is $P_{prec} = 3.8$ days. Interestingly, the shape 
of the 
optical light curves was found to be 
modulated on a $\sim 4 $ day period (Smale et al 1989). 
Homer et al (2000) have extensive X-ray and optical data
which seem 
consistent with our interpretation and with warping
of the disc.
Nodal precession of a warped disc may introduce futher
observable periodic modulations, analogous to the 35 day period
in Her X-1 (Scott \& Leahy 1999), and may be responsible for the
199 day cycle in 4U 1915-05 reported by Smale (1994).

\subsection{GX 9+9 (4U \thinspace 1728-16, Oph X-1, 2S \thinspace 1728-169)}

This system was suggested as a possible persistent superhumper
by Haswell and Abbott (1994) who found a $4.1744 \pm 0.0002 $ hr
modulation in I band photometry. Schaefer (1990) had previously
reported a B band modulation of period $ 4.198 \pm 0.028 $ hr,
and Hertz and Wood (1988) found an X-ray period of $4.19 \pm 0.02 $ hr.
These periods are clearly all mutually consistent.
With
$P_{\rm orb}=  4.2$ hrs and a neutron star primary, the system
has a mass ratio $q \la 0.28$ (Schaefer 1990) satisfying (1),
since the secondary mass $M_2$ cannot exceed the mass $\sim 0.5\msun$
of a main--sequence star filling the Roche lobe (see below).
We should therefore
expect GX 9+9 to harbour a persistently precessing elliptical disc.
The $\sim 10\%$ fractional amplitude of the optical modulations
in GX 9+9 is similar to those reported by O'Donoghue and Charles (1996)
for the superhumps in SXTs,
and the changes in the shapes of the optical light curves shown
in Haswell and Abbott (1994), figure 3, are suggestive of a light curve
shape 
modulation at $P_{\rm prec}$.

Since there are extensive RXTE observations of this source in hand,
a more precise X-ray period is likely to soon be available.  Clearly
this, and more extensive optical photometry is required to test
our hypothesis that GX 9+9 is an irradiated disc persistent superhumper.
Kong et al (2000) will address this.

\subsection{Many LMXB superhumpers?}
Table 1 lists LMXBs in order of increasing $P_{\rm orb}$
as reported in the literature.  The third column in this
table gives the nature of the modulation leading to the
$P_{\rm orb}$ determination.  For the eclipsing systems
and those in which orbital motion has been measured
(denoted either ``opt RV'' 
for mass donor radial velocity modulations
or ``pulsation RV''
for pulse timing
modulations)
$P_{\rm orb}$ is securely determined.  
The remaining 17 period determinations arise from modulations
in X-ray, UV, or optical flux and cannot be
identified as $P_{\rm orb}$ with certainty; these 17 periods
are in boldface.

We can show that the condition (1) for superhumps is likely to hold for
the first 11 systems listed in Table 1. The accretor presumably cannot be
less massive than a neutron star, and indeed is known to be such a star in
the first 10 cases because of the presence of X--ray bursts. Accordingly we can
assume that $M_1 \ga 1.4\msun$. The likely candidates for the donors
are either main--sequence stars, satisfying $M_2 \simeq 0.11P_{\rm
hr}\msun$ (e.g. King, 1988), 
or degenerate stars, obeying $M_2 \simeq 0.015(1 + X)^{5/2}P_{\rm
hr}^{-1}\msun$ (King, 1988) where $X$ is the fractional hydrogen content
by mass. (Degenerate companions are likely in the 5 systems in Table 1
with $P_{\rm hr} < 1$.) We thus find 
\begin{equation}
q_{\rm MS} \la 0.079P_{\rm hr},\ q_{\rm deg} \la 0.0115(1 + X)^{5/2}P_{\rm
hr}^{-1},
\label{q}
\end{equation}
for the two cases, showing that indeed the condition (1) for superhumps
holds for $P_{\rm orb} \la 4.2$~hr. The only possible exception to this
would be a donor star which has undergone thermal--timescale mass
transfer, as is thought to have occurred in for example Cygnus X--2 (King
\& Ritter, 1999). These stars are stripped down to their denser central
regions, and have larger masses when filling the Roche lobe at a given
period. However it is unlikely that periods $\la 2$~hr are
accessible to this kind of evolution (King et al., 2000).

Accordingly we expect that for the lower inclination systems
with $P_{\rm orb} \la 4.2$~hr, an optical or UV modulation is
more likely to be a superhump than an orbital modulation.
For high inclination systems satisfying (1) it is possible that
the X-ray dip behaviour depends on the disc's precessional and
superhump cycles, rather than straightforwardly indicating
$P_{\rm orb}$.

For a very extreme mass ratio, $q < 0.02$, the circularization
radius, 
$R_{\rm circ}$, is bigger than the 3:1 resonance radius (c.f
Frank et al. 1992, 
eqns (4.17, 5.75); Warner 1995 eqn (2.4a)).  In this case we expect
the disc to 
show persistent superhumps
for all values of
$\dot M.$
An LMXB with a $10\msun$ black hole primary and 
a secondary 
$\la 0.2\msun$
would satisfy this, and have $P_{\rm orb} \simeq 2$~hr.


\begin{table}
\begin{center}
\caption{Properties of LMXBs, adapted from Charles (1998), van Paradijs(1995) and van Paradijs (1998)}

\begin{tabular}{lcll} 
\hline
{\em Source} &  Period & Nature of &
X-ray type \\
 & (hrs) & modulation &\\ 
\hline

X1820-303 & {\bf 0.19} & X-ray & Burster, glob.cl. \\
4U~1850--087   & {\bf 0.34}  & UV &Burster, glob.cl. \\
X1626-673 & {\bf 0.7}  & opt sideband & Burster, Pulsar \\
X1832-330 & {\bf 0.73} & UV &Burster, glob.cl. \\
X1916-053 & {\bf 0.83} & X-ray, opt & Burster, Dipper \\
J1808.4-3658 & 2.0 & pulsation RV & Burster, Pulsar, Transient \\
X1323-619 & {\bf 2.9}  & X-ray dip & Burster, Dipper \\
X1636-536 & {\bf 3.8} & opt & Burster\\
X0748-676 & 3.8 & eclipsing & Burster, Dipper, Transient \\
X1254-690 & {\bf 3.9} & X-ray dip & Burster, Dipper \\
X1728-169 & {\bf 4.2} & opt &        \\
X1755-338 & {\bf 4.4} & X-ray dip & Dipper  \\
X1735-444 & {\bf 4.6} & opt & Burster \\
J0422+32 & 5.1 & opt RV & BH,  Transient \\
X2129+470 & {\bf 5.2} & opt & ADC     \\
X1822-371 & 5.6 & eclipsing & ADC     \\
J2123-058 & 6.0 & eclipsing & Burster, Transient \\
N Vel 93 & 6.9 & opt RV & BH,  Transient \\
X1658-298 & {\bf 7.2} & X-ray dip & Burster, Dipper \\
A0620-00 & 7.8 & opt RV & BH,  Transient \\
G2000+25 & 8.3 & opt RV & BH,  Transient \\
A1742-289 & 8.4 & eclipsing & Burster, Transient \\
X1957+115 & {\bf 9.3} & opt &         \\
N Mus 91 & 10.4 & opt RV & BH,  Transient \\
N Oph 77 & 12.5 & opt RV & BH,  Transient \\
Cen X-4 & 15.1 & opt RV & Burster, Transient \\
X2127+119 & 17.1 & eclipsing & Burster, ADC, glob.cl. \\
Aql X-1 & {\bf 19} & opt & Burster, Transient \\
Sco X-1   & {\bf 19.2} & opt & Prototype LMXB \\
X1624-490 & {\bf 21} & X-ray dip & Dipper \\
N Sco 94 & 62.6 & opt RV & BH,  Transient \\
V404 Cyg & 155.4 & opt RV & BH,  Transient \\
2S0921-630 & 216 & eclipsing & ADC \\
Cyg X-2 & 235 & opt RV & Burster \\
J1744-28  & 283   & pulsation RV & Burster, Pulsar, Transient \\

\hline
\end{tabular}

\end{center}
\end{table}

If it were possible to 
show unambiguously
that an LMXB with
$P_{\rm orb} > 5-7~{\rm hr}$ exhibited superhumps,
this would suggest it harboured a black
hole primary.
We note, of course, that superhumps alone cannot
provide proof of a black hole primary, 
as the secondary could 
have lower mass
than a main sequence star,
for example if it were slightly evolved. This might allow superhumps even
though the primary was a neutron star.
However, coupled with a lack of  Type I X-ray bursts, 
and with the presence of X-ray spectral and timing indicators of black hole
candidacy, this line of argument could prove useful,
since it it extremely difficult to
obtain a dynamical mass determination for the accretor in
a persistent
LMXB.  It is notable that {\bf all} the identified black holes 
among LMXB primaries (denoted BH in Table 1) are in
transient systems.  The nature of the compact objects
in  steady LMXBs which neither pulse nor burst
is currently an undesirably open question.

\section*{Acknowledgments}

This work is supported by grant F/00-180/A 
from the Leverhulme Trust (held jointly by PAC, CAH and ARK). 
ARK gratefully acknowledges a PPARC Senior Fellowship.

\label{lastpage}

\end{document}